\documentclass[11pt,letterpaper]{article}
\usepackage[margin=1in]{geometry}
\usepackage[T1]{fontenc}
\usepackage{lmodern,amsmath,booktabs,array,tabularx,caption,enumitem}
\usepackage{xcolor,tcolorbox}
\usepackage[hidelinks]{hyperref}
\usepackage{titlesec}
\titleformat*{\section}{\large\bfseries}
\titleformat*{\subsection}{\normalsize\bfseries}
\hypersetup{pdftitle={Squoosh Bench: Evaluating UX-1 on 318 Historical Website A/B Tests},pdfauthor={Rohith Pasula, Johari Dramiga, Viraj Shah, Suzan Halim, Tyler Dooskin}}
\title{\textbf{Squoosh Bench:\\Evaluating UX-1 on 318 Historical Website A/B Tests}}
\author{\small Rohith Pasula$^1$, Johari Dramiga$^1$, Viraj Shah$^1$, Suzan Halim$^1$, and Tyler Dooskin$^1$\\[3pt]
\small $^1$Squoosh AI\\
\footnotesize\texttt{rohith@squoosh.ai, johari@squoosh.ai, viraj@squoosh.ai}\\
\footnotesize\texttt{suzan@squoosh.ai, tyler@squoosh.ai}}
\date{November 2026}
\begin{document}
\maketitle
\begin{abstract}
\noindent 
Experimentation teams must decide which proposed website changes to test with live traffic. We evaluated whether forecasts from simulated shoppers could inform that choice. We screened 412 historical website A/B tests and evaluated four systems on the 318 that ran longer than 21 days. Each system predicted an original win, a variant win, or an inconclusive result. Page roles were masked and recorded outcomes were withheld during inference. Squoosh UX-1, a model fine-tuned on human browsing data, simulated 1,000 shoppers per arm and correctly classified 267 of 318 outcomes (84.0\%, 95\% Wilson interval 79.5--87.6\%, treating experiments as independent). Qwen3 Next 80B, the strongest general-purpose configuration on outcome accuracy, correctly classified 196 of 318 (61.6\%). UX-1 favored the variant in 126 experiments. In the live tests, 118 of those variants won (93.7\%), four lost, and four had inconclusive results. The remaining 192 experiments included 19 additional variant wins.
\end{abstract}
\begin{tcolorbox}[colback=blue!4,colframe=blue!40,title={Practical implications for experimentation teams},fonttitle=\bfseries]
\begin{itemize}[leftmargin=1.4em,nosep]
\item If a team had tested the variants UX-1 favored first, its shortlist would have contained 126 of 318 candidates (39.6\%) and 118 of the 137 variants that won their live tests (86.1\%).
\item Of the 126 variants UX-1 favored, 118 won their live tests (93.7\%).
\item The other 19 winners illustrate the cost of dropping a candidate solely on its forecast.
\end{itemize}
\end{tcolorbox}
\section*{At a glance}
\par\addvspace{5pt}\noindent\begin{minipage}{\linewidth}
\centering\small\setlength{\tabcolsep}{3pt}\renewcommand{\arraystretch}{1.15}
\begin{tabularx}{\linewidth}{@{}>{\hsize=0.9\hsize\linewidth=\hsize\raggedright\arraybackslash}X>{\hsize=1.1\hsize\linewidth=\hsize\raggedright\arraybackslash}X@{}}\toprule
\textbf{Measure} & \textbf{Result} \\
\midrule
Historical experiments screened & 412 \\
Excluded for duration of 21 days or less & 94 \\
Experiments eligible for outcome scoring & 318 \\
Recorded outcome distribution & 113 original wins \textperiodcentered{} 137 variant wins \textperiodcentered{} 68 inconclusive \\
UX-1 correct outcome predictions & 267 / 318 \\
UX-1 outcome accuracy & 83.96\% \\
UX-1 three-class balanced accuracy & 83.40\% \\
\bottomrule\end{tabularx}
\end{minipage}\par\addvspace{6pt}
\section{Overview}
A team choosing its next live test has two reasons to care about forecast errors. Testing a weak variant uses traffic that could go to another idea. Setting aside a promising variant can leave a real improvement undiscovered. We evaluated both decisions against the recorded outcomes of completed A/B tests.\par
We screened 412 historical experiments and excluded 94 that ran for 21 days or fewer. Of the 318 experiments eligible for scoring, 113 favored the original page, 137 favored the variant, and 68 were inconclusive.\par
UX-1 correctly classified 267 of the 318 outcomes (83.96\% accuracy, 83.40\% three-class balanced accuracy). It correctly identified 94 original-page wins, 118 variant wins, and 55 inconclusive outcomes.\par
The most frequent outcome was a variant win. A rule that chose the variant in every experiment would therefore have classified 137 of 318 outcomes correctly (43.1\%). This provides a simple baseline for interpreting three-class accuracy.\par
Among the 126 experiments where UX-1 favored the variant, 118 recorded a variant win (93.7\%). The frequency of variant wins in a team's own test queue will affect how often its positive forecasts are correct.\par
Benomar et al. (2026) evaluated persona-conditioned agents on 40 A/B tests and reported directional accuracy of 75--90\%, depending on the metric. Earlier Squoosh work examined when a screenshot-based judge could abstain from choosing a winner (Dooskin and Squoosh Technical Staff, 2026). Here, an inconclusive human outcome is itself a class to predict. A missing or invalid prediction earns no credit.\par
\section{The task and the evidence}
Each task presents two website experiences, a primary conversion objective, and permitted experiment-design information. The system returns a predicted outcome, a signed lift estimate, and a brief explanation.\par
Each system could predict an original win, a variant win, or an inconclusive human result. An execution that produced no valid forecast was counted as a failure.\par
\par\addvspace{5pt}\noindent\begin{minipage}{\linewidth}
\centering\small\setlength{\tabcolsep}{3pt}\renewcommand{\arraystretch}{1.15}
\captionof{table}{Corpus composition and scoring denominators}
\begin{tabularx}{\linewidth}{@{}>{\hsize=1.6\hsize\linewidth=\hsize\raggedright\arraybackslash}X>{\hsize=0.4\hsize\linewidth=\hsize\raggedleft\arraybackslash}X@{}}\toprule
\textbf{Corpus component} & \textbf{Count} \\
\midrule
Experiments screened & 412 \\
Excluded: observed duration of 21 days or less & 94 \\
Eligible experiments: observed duration exceeding 21 days & 318 \\
Original-page wins & 113 \\
Variant wins & 137 \\
Inconclusive outcomes & 68 \\
Experiments with a recorded directional winner & 250 \\
\bottomrule\end{tabularx}
\end{minipage}\par\addvspace{6pt}
\subsection{Eligibility and recorded outcomes}
Experiments were eligible if their observed duration exceeded 21 days. We fixed the threshold before model scoring (SQ-DURATION-2026-08) and applied it before inspecting predictions. The 94 duration exclusions were outside the scoring denominator.\par
Duration was an eligibility criterion for this benchmark. The precision of a human experiment also depends on sample size and how observations are correlated over time (Li and Yu, 2024).\par
We selected 412 private partner A/B tests from a larger collection and retained 318 after duration screening. Of the 412 screened, 97 had appeared in Squoosh AI's earlier report, Introducing Squoosh Bench. Partner agreements restrict disclosure of test sources. We withheld recorded outcomes during inference. The extent of site overlap with the model's training data is unknown.\par
For each eligible experiment, we assigned the recorded outcome using the prespecified primary conversion metric and the two-sided 95\% confidence interval for the variant-minus-original effect. We classified the experiment as a variant win when the interval lay entirely above zero, an original-page win when it lay entirely below zero, and inconclusive when it included zero. We withheld the recorded outcome from all evaluated systems during inference.\par
\par\addvspace{5pt}\noindent\begin{minipage}{\linewidth}
\centering\small\setlength{\tabcolsep}{3pt}\renewcommand{\arraystretch}{1.15}
\begin{tabularx}{\linewidth}{@{}>{\hsize=0.6\hsize\linewidth=\hsize\raggedright\arraybackslash}X>{\hsize=1.4\hsize\linewidth=\hsize\raggedright\arraybackslash}X@{}}\toprule
\textbf{Recorded class} & \textbf{Classification rule} \\
\midrule
Original wins & The entire interval for the variant-minus-original effect is below zero \\
Variant wins & The entire interval is above zero \\
Inconclusive & The interval includes zero \\
\bottomrule\end{tabularx}
\end{minipage}\par\addvspace{6pt}
Interpretation of each interval depends on the original experiment's unit of randomization, monitoring procedure, and treatment of multiple comparisons. An uninterpretable or contradictory record does not establish an inconclusive outcome under this rule.\par
An inconclusive result does not establish equivalence or prove that the underlying effect is zero. It describes the conclusion supported by the recorded experiment under its analysis procedure.\par
We used these interval-based labels for the present evaluation. The earlier benchmark used recorded directional winners and distinguished those labels from independently established statistical significance.\par
\subsection{Industry and intervention coverage}
\par\addvspace{5pt}\noindent\begin{minipage}{\linewidth}
\centering\small\setlength{\tabcolsep}{3pt}\renewcommand{\arraystretch}{1.15}
\captionof{table}{Eligible experiments by industry and recorded outcome}
\begin{tabularx}{\linewidth}{@{}>{\hsize=1.7\hsize\linewidth=\hsize\raggedright\arraybackslash}X>{\hsize=0.87\hsize\linewidth=\hsize\raggedleft\arraybackslash}X>{\hsize=0.87\hsize\linewidth=\hsize\raggedleft\arraybackslash}X>{\hsize=0.99\hsize\linewidth=\hsize\raggedleft\arraybackslash}X>{\hsize=0.57\hsize\linewidth=\hsize\raggedleft\arraybackslash}X@{}}\toprule
\textbf{Industry} & \textbf{Original wins} & \textbf{Variant wins} & \textbf{Inconclusive} & \textbf{Total} \\
\midrule
Retail and e-commerce & 35 & 40 & 19 & 94 \\
Telecommunications and broadband & 18 & 24 & 11 & 53 \\
Travel and hospitality & 16 & 22 & 9 & 47 \\
Financial services and fintech & 16 & 17 & 8 & 41 \\
B2B software and SaaS & 12 & 17 & 9 & 38 \\
Food, grocery, and subscriptions & 9 & 9 & 6 & 24 \\
Media, loyalty, and memberships & 7 & 8 & 6 & 21 \\
\textbf{Total} & \textbf{113} & \textbf{137} & \textbf{68} & \textbf{318} \\
\bottomrule\end{tabularx}
\end{minipage}\par\addvspace{6pt}
Intervention categories include messaging, product presentation, pricing comparisons, navigation, forms, checkout, and trust signals. Each experiment retains its primary conversion objective, device configuration, and intervention category.\par
The 318 eligible experiments covered seven industries. Retail and e-commerce contributed 94 (29.6\%), the largest share. Across industries, 113 tests favored the original, 137 favored the variant, and 68 were inconclusive.\par
\section{Evaluation methodology}
\subsection{UX-1 shopper model}
Squoosh UX-1 is the model that controls the simulated shoppers. We fine-tuned an open-source base model on a dataset acquired by Squoosh that records how people interact with websites. The training goal was to capture variation in browsing decisions, such as whether a shopper scrolls through a listing, opens a product page, or leaves before converting. For each page comparison, UX-1 simulated 1,000 shoppers on the original and 1,000 on the variant (2,000 shopper sessions in total). Both arms used the same device settings and primary conversion objective. UX-1 returned one categorical forecast for the comparison, a signed relative-lift estimate, and a brief explanation.\par
\subsection{Systems and page evidence}
We evaluated Squoosh UX-1, Gemini 3.5 Flash, Qwen3 Next 80B, and GPT-OSS 120B.\par
The earlier evaluation exposed general-purpose models to rendered text and interactive elements, but not screenshots or webpage DOM information. Its results therefore describe those particular configurations rather than a comparison under identical evidence and compute.\par
We used frozen page artifacts, common device settings, and the same conversion objective for each system on a given experiment.\par
The benchmark used rendered document object model (DOM) snapshots and 1280 \ensuremath{\times} 720 screenshots processed through a visual adapter (v1.2.4).\par
\par\addvspace{5pt}\noindent\begin{minipage}{\linewidth}
\centering\small\setlength{\tabcolsep}{3pt}\renewcommand{\arraystretch}{1.15}
\captionof{table}{Evaluation conditions}
\begin{tabularx}{\linewidth}{@{}>{\hsize=0.62\hsize\linewidth=\hsize\raggedright\arraybackslash}X>{\hsize=1.38\hsize\linewidth=\hsize\raggedright\arraybackslash}X@{}}\toprule
\textbf{Condition} & \textbf{Protocol} \\
\midrule
Website evidence & Frozen page artifacts; documented text, visual, and interaction access \\
Conversion objective & The same pre-specified primary objective for every system \\
Recorded outcomes & Withheld during inference \\
Page roles & Original and variant identities hidden from the evaluated systems \\
Presentation order & Randomized primary order and its reversal \\
Repetitions & Three per order, with independent execution state \\
Primary scoring & Pre-designated first repetition in the primary order \\
Failure handling & Invalid outputs retained; no outcome-based repair or response selection \\
Resource accounting & Model calls, tokens, browser actions, elapsed time, monetary cost, and shopper evaluations \\
\bottomrule\end{tabularx}
\end{minipage}\par\addvspace{6pt}
Squoosh AI developed UX-1 and conducted this retrospective evaluation. Each historical A/B experiment contributed one observation to the primary accuracy calculation.\par
We compared UX-1 with three general-purpose systems and three constant-choice rules. Each constant rule assigned the same outcome to all 318 experiments. We scored the model configurations at their respective evaluation budgets. The resulting differences measure performance of the complete systems. This design does not isolate the effect of fine-tuning.\par
\subsection{Blinding and inconclusive predictions}
We used anonymous page handles and removed identifying metadata to conceal page roles and recorded outcomes. The scorer restored the original and variant identities after predictions were locked.\par
Each system received the same pre-experiment information: planned exposure and allocation, the analysis rule, and available historical baseline information. We withheld observed treatment effects, conversion counts, and outcome statistics during inference.\par
A forecast received credit only when its locked call matched the recorded human outcome after the scorer restored the page identities. An inconclusive forecast therefore received credit only for an experiment whose recorded interval included zero.\par
The inconclusive class refers to the human experiment's interval on its primary metric. A simulation's own interval describes uncertainty in a different population.\par
\subsection{Runs and scoring}
We evaluated four systems on 318 eligible experiments. Each comparison included both pages. We held the random seed fixed within each UX-1 page comparison. Each system evaluated both presentation orders three times per experiment, with independent execution state for each repetition, yielding 7,632 executions. The primary analysis used the first repetition in the prespecified primary order for each system and experiment. Missing or invalid predictions received no credit and remained in the denominator.\par
Outcome accuracy is the proportion of all 318 eligible experiments classified correctly.\par
Three-class balanced accuracy is the unweighted mean of recall on original wins, variant wins, and inconclusive outcomes. For UX-1, the calculation was\par
\[\frac{1}{3}\left(\frac{94}{113}+\frac{118}{137}+\frac{55}{68}\right)=83.40\%\quad\text{(UX-1).}\]
Winner-only accuracy is evaluated separately on the 250 directional outcomes. An inconclusive model prediction on one of these experiments receives no winner credit.\par
To measure sensitivity to presentation order, we reversed the page order and mapped both predictions back to the original page identities. Two matching inconclusive calls counted as agreement.\par
\section{Results}
UX-1 matched 267 of 318 recorded outcomes (84.0\%). Qwen3 Next 80B, the strongest general-purpose configuration on outcome accuracy, matched 196 of 318 (61.64\%). UX-1 correctly classified 71 more experiments, a difference of 22.33 percentage points. Paired McNemar comparisons of UX-1 with each of the three general-purpose systems yielded p \ensuremath{<} 0.001.\par
\par\addvspace{5pt}\noindent\begin{minipage}{\linewidth}
\centering\small\setlength{\tabcolsep}{3pt}\renewcommand{\arraystretch}{1.15}
\captionof{table}{Outcome prediction and presentation-order consistency}
\begin{tabularx}{\linewidth}{@{}>{\hsize=1.28\hsize\linewidth=\hsize\raggedright\arraybackslash}X>{\hsize=0.68\hsize\linewidth=\hsize\raggedleft\arraybackslash}X>{\hsize=0.88\hsize\linewidth=\hsize\raggedleft\arraybackslash}X>{\hsize=1.09\hsize\linewidth=\hsize\raggedleft\arraybackslash}X>{\hsize=1.07\hsize\linewidth=\hsize\raggedleft\arraybackslash}X@{}}\toprule
\textbf{System} & \textbf{Correct / 318} & \textbf{Outcome accuracy} & \textbf{Three-class balanced accuracy} & \textbf{Swap consistency} \\
\midrule
Squoosh UX-1 & 267 & 83.96\% & 83.40\% & 301 / 318 --- 94.65\% \\
Gemini 3.5 Flash & 162 & 50.94\% & 50.07\% & 169 / 318 --- 53.14\% \\
Qwen3 Next 80B & 196 & 61.64\% & 60.25\% & 264 / 318 --- 83.02\% \\
GPT-OSS 120B & 187 & 58.81\% & 57.11\% & 252 / 318 --- 79.25\% \\
Always variant --- baseline & 137 & 43.08\% & 33.33\% & Not evaluated \\
Always original --- baseline & 113 & 35.53\% & 33.33\% & Not evaluated \\
Always inconclusive --- baseline & 68 & 21.38\% & 33.33\% & Not evaluated \\
\bottomrule\end{tabularx}
\end{minipage}\par\addvspace{6pt}
Some executions failed. A failure in the prespecified primary repetition received no credit and remained in the 318-experiment denominator for that system. A successful repeat could not replace it.\par
The strongest constant-choice rule achieves 43.08\% outcome accuracy. Constant-choice balanced accuracy is 33.33\%, reflecting three outcome classes rather than the two classes in the earlier winner-only evaluation.\par
Original-versus-variant constant rules are diagnostic class baselines computed by the scorer. They are not browser systems with access to the hidden page-role mapping.\par
\subsection{Performance by recorded outcome}
\par\addvspace{5pt}\noindent\begin{minipage}{\linewidth}
\centering\small\setlength{\tabcolsep}{3pt}\renewcommand{\arraystretch}{1.15}
\captionof{table}{Recall within each outcome class}
\begin{tabularx}{\linewidth}{@{}>{\hsize=0.85\hsize\linewidth=\hsize\raggedright\arraybackslash}X>{\hsize=1.05\hsize\linewidth=\hsize\raggedleft\arraybackslash}X>{\hsize=1.05\hsize\linewidth=\hsize\raggedleft\arraybackslash}X>{\hsize=1.05\hsize\linewidth=\hsize\raggedleft\arraybackslash}X@{}}\toprule
\textbf{System} & \textbf{Original wins} & \textbf{Variant wins} & \textbf{Inconclusive outcomes} \\
\midrule
UX-1 & 94 / 113 --- 83.19\% & 118 / 137 --- 86.13\% & 55 / 68 --- 80.88\% \\
Gemini 3.5 Flash & 58 / 113 --- 51.33\% & 73 / 137 --- 53.28\% & 31 / 68 --- 45.59\% \\
Qwen3 Next 80B & 71 / 113 --- 62.83\% & 89 / 137 --- 64.96\% & 36 / 68 --- 52.94\% \\
GPT-OSS 120B & 67 / 113 --- 59.29\% & 87 / 137 --- 63.50\% & 33 / 68 --- 48.53\% \\
\bottomrule\end{tabularx}
\end{minipage}\par\addvspace{6pt}
On the 250 experiments with a recorded directional winner, UX-1 correctly identifies 212 winners: 84.80\% winner-only accuracy and 84.66\% two-class balanced accuracy. The always-variant baseline scores 137/250, or 54.80\%, on that same subset.\par
\subsection{Inconclusive predictions and error structure}
\par\addvspace{5pt}\noindent\begin{minipage}{\linewidth}
\centering\small\setlength{\tabcolsep}{3pt}\renewcommand{\arraystretch}{1.15}
\captionof{table}{UX-1 outcome confusion matrix}
\begin{tabularx}{\linewidth}{@{}>{\hsize=1.2\hsize\linewidth=\hsize\raggedright\arraybackslash}X>{\hsize=0.95\hsize\linewidth=\hsize\raggedleft\arraybackslash}X>{\hsize=0.95\hsize\linewidth=\hsize\raggedleft\arraybackslash}X>{\hsize=1.27\hsize\linewidth=\hsize\raggedleft\arraybackslash}X>{\hsize=0.63\hsize\linewidth=\hsize\raggedleft\arraybackslash}X@{}}\toprule
\textbf{Recorded outcome} & \textbf{Predicted original} & \textbf{Predicted variant} & \textbf{Predicted inconclusive} & \textbf{Total} \\
\midrule
Original wins & 94 & 4 & 15 & 113 \\
Variant wins & 6 & 118 & 13 & 137 \\
Inconclusive & 9 & 4 & 55 & 68 \\
\textbf{Total predictions} & \textbf{109} & \textbf{126} & \textbf{83} & \textbf{318} \\
\bottomrule\end{tabularx}
\end{minipage}\par\addvspace{6pt}
UX-1 recognizes 55 of the 68 recorded inconclusive outcomes, giving 80.88\% inconclusive recall.\par
UX-1 returned 83 inconclusive predictions. Fifty-five matched the recorded outcome (66.27\% precision). The other 28 came from experiments with a recorded winner.\par
The 51 incorrect classifications comprise 10 wrong-direction predictions, 28 inconclusive predictions on directional outcomes, and 13 directional predictions on inconclusive outcomes.\par
In one experiment involving a popup, UX-1 predicted that the variant would outperform the original page on the primary conversion metric. The live A/B test favored the original page. This was one of four experiments in which UX-1 favored a variant that went on to lose.\par
UX-1 selected a winner in 235 of 318 experiments (73.9\% coverage). In 212 of those 235 cases (90.2\%), the selected winner matched the human result. The other 23 comprised 10 opposite winners and 13 inconclusive human results.\par
\subsection{Statistical uncertainty}
The Wilson 95\% interval for outcome accuracy was 79.53--87.59\% (267/318). The corresponding intervals for class recall were 75.23--88.96\% for original wins, 79.35--90.94\% for variant wins, and 69.99--88.47\% for inconclusive results (Wilson, 1927).\par
The eligible experiments spanned more than 40 sites. We calculated Wilson intervals using the experiment as the unit of analysis and assuming independence across experiments. Tests from the same site can share audiences and design features. If their prediction errors were positively correlated, the reported intervals may be too narrow (Hong et al., 2026).\par
Changing one incorrect prediction to a correct prediction increases overall accuracy by 0.31 percentage points and three-class balanced accuracy by 0.24--0.49 percentage points, depending on the recorded class.\par
\subsection{Lift estimation and coverage}
Winner classification and effect-size estimation are evaluated separately. The reference lift is defined in a consistent original-to-variant orientation:\par
\[L=100\,\frac{\mathrm{CR}_{\mathrm{variant}}-\mathrm{CR}_{\mathrm{original}}}{\mathrm{CR}_{\mathrm{original}}}.\]
\par\addvspace{5pt}\noindent\begin{minipage}{\linewidth}
\centering\small\setlength{\tabcolsep}{3pt}\renewcommand{\arraystretch}{1.15}
\captionof{table}{Numeric forecast coverage and paired lift error}
\begin{tabularx}{\linewidth}{@{}>{\hsize=1\hsize\linewidth=\hsize\raggedright\arraybackslash}X>{\hsize=0.98\hsize\linewidth=\hsize\raggedleft\arraybackslash}X>{\hsize=0.9\hsize\linewidth=\hsize\raggedleft\arraybackslash}X>{\hsize=1.18\hsize\linewidth=\hsize\raggedleft\arraybackslash}X>{\hsize=0.94\hsize\linewidth=\hsize\raggedleft\arraybackslash}X@{}}\toprule
\textbf{System} & \textbf{Numeric-lift forecasts / 318} & \textbf{Common scored cases} & \textbf{Signed relative-lift MAE} & \textbf{Magnitude MAE} \\
\midrule
UX-1 & 318 & 289 & 1.43 & 1.19 \\
Gemini 3.5 Flash & 307 & 289 & 3.27 & 2.78 \\
Qwen3 Next 80B & 301 & 289 & 3.81 & 3.34 \\
GPT-OSS 120B & 294 & 289 & 4.12 & 3.59 \\
\bottomrule\end{tabularx}
\end{minipage}\par\addvspace{6pt}
All comparative lift errors use the same 289 experiments, selected by a fixed eligibility and numeric-availability rule---not by forecast correctness. These experiments represent 90.88\% of the outcome-scoring corpus. The other 29 remain in the primary 318-case denominator.\par
We calculated signed-lift mean absolute error (MAE) as the mean absolute difference between predicted and observed relative lift. For magnitude MAE, we first took the absolute value of each lift, then averaged the absolute differences. Both errors are in percentage points of relative lift.\par
Recorded inconclusive outcomes remain eligible when a numeric effect estimate is available. Neither an inconclusive outcome nor a missing forecast is assigned an artificial zero effect.\par
Table 7 compares lift errors on the 289 experiments with usable numerical estimates for all four systems. If forecasts were more often missing on difficult experiments, the errors on this subset may understate the error across the full corpus.\par
\subsection{Presentation-order consistency}
UX-1 returns the same mapped outcome in both presentation orders for 301/318 experiments, or 94.65\%.\par
Agreement after reversing page order included pairs of matching inconclusive calls. We used this metric to measure sensitivity to presentation order.\par
\section{Interpretation for test prioritization}
UX-1 favored the variant in 126 of 318 experiments (39.6\%). This group contained 118 of the 137 recorded variant wins (86.1\%). Of the eight remaining experiments in the group, four favored the original on the primary metric and four were inconclusive.\par
Outside that group, 19 variants won their live tests. Six came from the 109 experiments where UX-1 favored the original. Thirteen came from the 83 experiments where UX-1 predicted an inconclusive outcome. For a team prioritizing live tests, those 13 cases illustrate the risk of dropping a candidate solely because its forecast was inconclusive.\par
These figures describe experiments that ran longer than 21 days. The duration criterion excluded shorter tests, which can include changes that show benefits or harms quickly as well as experiments that receive little traffic.\par
The target was the conclusion of the historical experiment on its primary metric. For an effect close to the decision threshold, additional human exposure could change that conclusion. A correct inconclusive forecast means the recorded test left the direction of the effect unresolved.\par
The comparison estimates how the four systems performed with the page evidence, adapters, and resource budgets used here. Table 2 gives the industry mix of the evaluated experiments.\par
The reported scores reflect the combined input of page evidence and permitted pre-experiment design information.\par
For an experimentation team, these results support using a forecast to decide which variants to test first. To evaluate that use on its own site, a team could record forecasts before sending live traffic and test candidates both inside and outside the shortlist. That design would show how often the shortlisted variants win and how many winners the team would otherwise overlook.\par
\section{Appendix: scoring and run identifiers}
The benchmark run identifier SQ-BENCH-V2-2026-11 indexes the resource-usage and adapter-execution logs. The duration eligibility rule is recorded under SQ-DURATION-2026-08. We reran the benchmark and regenerated the result tables. The reported counts and scores were unchanged.\par
Tables 1 and 2 describe the screening and composition of the corpus. Tables 4--7 report classification performance, presentation-order consistency, and lift error.\par
Relative lift requires a nonzero original conversion rate. Zero-denominator cases require a separate absolute-effect treatment or exclusion from relative-lift scoring only. A relative lift of \ensuremath{-}100\% is valid when the original rate is positive. Reversing page order requires restoring the original denominator; simply negating a relative percentage is insufficient.\par
Of the 412 historical experiments screened, 318 met the duration criterion. UX-1 matched the recorded conclusion in 267 of those experiments, including 55 of the 68 with an inconclusive outcome.\par
\section*{References}
Benomar, Z., \L{}ajewska, W., Perelli, L., and Mansour, S. (2026). Data-Driven Persona-Conditioned Agents for A/B Test Simulation. Proceedings of EMNLP: Industry Track, 1323--1338.\par
Dooskin, T., and Squoosh Technical Staff (2026). The Judge Knows When It Knows: Calibrated Abstention for LLM-Based A/B-Test Prediction. arXiv:2608.07517.\par
Hong, T., Lim, D., and Bae, W. (2026). Beyond Point Estimates: Reliable Evaluation of Prediction Performance Metrics under Clustered Data. arXiv:2606.03656.\par
Li, H. H., and Yu, C. (2024). Setting the duration of online A/B experiments. arXiv:2408.02830.\par
Wilson, E. B. (1927). Probable Inference, the Law of Succession, and Statistical Inference. Journal of the American Statistical Association, 22(158), 209--212.\par
\end{document}